\documentclass[preprint,pra,aps,showpacs,preprintnumbers,amsmath,amssymb,eqsecnum]{revtex4}

\usepackage{graphicx}
\usepackage{dcolumn}
\usepackage{bm}
\usepackage{verbatim}
\usepackage{epsfig}
\usepackage{epstopdf}

\newcommand\beq{\begin{equation}}
\newcommand\eeq{\end{equation}}
\newcommand\bea{\begin{eqnarray}}
\newcommand\eea{\end{eqnarray}}

\def\av{{\bf a}}
\def\bv{{\bf b}}

\def\half{\frac {1} {2}}

\def\x0{{{\bf x}_0}}

\def \MR {\rm MR}
\def\LG{\rm LG}
\def\NSIT{\rm NSIT}
\def\NIM{\rm NIM}
\def\Ind{\rm AoT}
\def\Ind{\rm Ind}

\begin{document}


\title{Comparing Conditions for Macrorealism:  Leggett-Garg Inequalities
vs No-Signaling in Time}

\author{J.J.Halliwell}%
\email{j.halliwell@imperial.ac.uk}

\affiliation{Blackett Laboratory \\ Imperial College \\ London SW7 2BZ \\ UK }



\begin{abstract}
We consider two different types of conditions which were proposed to test macrorealism in the context of 
a system described by a single dichotomic variable $Q$.
This is the view
that a macroscopic system evolving in time possesses definite properties which can be determined without disturbing the future or past state.
The Leggett-Garg (LG) inequalities, the most commonly-studied test, are only necessary conditions for macrorealism, but building on earlier work (Phys.Rev. A93, 022123 (2016)) it is shown that when the four three-time LG inequalities are augmented with a certain set of two-time inequalities also of the LG form, Fine's theorem applies and these augmented conditions are then both necessary and sufficient. A comparison is carried out with a very different set of necessary and sufficient conditions for macrorealism, namely the no-signaling in time (NSIT) conditions proposed by Brukner, Clemente, Kofler and others, which ensure that all probabilities for $Q$ at one and two times are independent of whether earlier or intermediate measurements are made in a given run, and do not require (but  imply) the LG inequalities. We argue that tests based on the LG inequalities have the form of very weak classicality conditions and can be satisfied in the face of moderate interference effects, 
but those based on NSIT conditions have the form of much stronger coherence witness conditions, satisfied only for zero interference. The two tests differ in their implementation of non-invasive measurability so are testing different notions of macrorealism:
the augmented LG tests are indirect, entailing a combination of the results of different experiments with only compatible  
quantities measured in each experimental run, in close analogy with Bell tests, and
are primarily tests for macrorealism {\it per se}; by contrast the NSIT tests entail sequential measurements of incompatible quantities and are primarily tests for non-invasiveness.
\end{abstract}

\pacs{03.65.Ta, 03.65.Ud, 03.65.Yz}








\maketitle

\section{Introduction}

\subsection{Macrorealism and the Leggett-Garg Inequalities}

The notion of macroscopic realism (macrorealism), introduced by Leggett and Garg \cite{LG1,L1,ELN}, is the idea that a time-evolving macroscopic system can possess definite properties at a number of times uninfluenced by measurements of it. 
Macrorealism (MR) was proposed by way of analogy to the notion of local realism for spatially entangled systems and indeed leads to a set of inequalities obeyed by the temporal correlation functions of a single system, similar to the Bell and CHSH inequalities.
Most investigations to date focus on a single dichomotic variable $Q$ which is measured in various ways at three (or more) times leading to the determination of the temporal correlation functions of the form,
\beq
C_{12} = \langle Q(t_1) Q(t_2) \rangle.
\label{corr}
\eeq
These are argued, for a macrorealistic theory, to obey the Leggett-Garg (LG) inequalities,
\bea
1 + C_{12} + C_{23} + C_{13} & \ge & 0,
\label{LG1}
\\
1 - C_{12} - C_{23} + C_{13} & \ge & 0,
\label{LG2}
\\
1 + C_{12} - C_{23} - C_{13} & \ge & 0,
\\
1 - C_{12} + C_{23} - C_{13} & \ge & 0,
\label{LG4}
\eea
which are identical in mathematical form to the Bell inequalities. Measurements at four times lead to a set of eight LG inequalities identical in mathematical form to the CHSH inequalities.

To derive these inequalities, the notion of macrorealism is broken down into three separate assumptions. These are:

\noindent{\bf 1.} Macrorealism {\it per se} (MRps): the system is in one of the states available to it at each moment of time.

\noindent{\bf 2.} Non-invasive measurability (NIM): it is possible in principle to determine the state of the system without disturbing the subsequent dynamics.

\noindent{\bf 3.} Induction (Ind): future measurements cannot affect the present state.

Any experimental test thus tests the combination of these assumptions. Induction is always taken for granted so what is being tested is the combination of MRps and NIM. To ensure NIM, Leggett and Garg proposed that the measurement of the correlation functions be carried out using ideal negative measurements, in which the detector is coupled to, say, only the $Q=+1$ state, at the first time, and a null result then permits us to deduce that the system is in the $Q=-1$ state but without any interaction taking place, from the macrorealistic perspective. This procedure rules out alternative classical explanations of the correlation functions \cite{Mon,Ye1,Guh} analogous to the way in which signaling is ruled out in Bell experiments and has been successfully implemented in a number of recent experiments \cite {Knee,Rob,KBLL,EmINRM}. Many other experimental tests of the LG inequalities have also been carried out, on a variety of different physical systems \cite{LGexpt,SimpLG}.

Numerous aspects of the LG inequalities and the question of what they actually test for have been significantly clarified by Maroney and Timpson \cite{MaTi}. They argued that MRps actually comes in three different varieties. 
The first, which they refer to as ``operational eigenstate mixture macrorealism'', includes spontaneous collapse models, such as those of the Ghirardi-Rimini-Weber type \cite{GRW}. This is the notion of MRps that Leggett and Garg seemed to allude to in their early papers and it is only this type of MRps that can be ruled out by the LG inequalities, so this is the variety of MRps we have in mind in the present paper.
Of the other two varieties of MRps,
the most significant one is realist theories of the Bohmian type, which cannot be ruled out by the LG inequalities, unless some sort of locality arguments can be brought to bear on the experimental arrangement. (This was also noted by Bacciagalupi \cite{Bac}).
The remaining type have the form of restricted Bohmian theories (and include the Kochen-Specker model \cite{KS} for two-dimensional Hilbert spaces) but for Hilbert spaces of dimension greater than $3$ these can be ruled out \cite{Ma2}.

The NIM requirement is the source of considerable debate in LG inequality tests. Part of the debate is that, like MRps, NIM 
can be interpreted in a number of different ways depending on exactly what is measured and how the measurements are carried out.
Most experimental tests of the LG inequalities measure the three correlation functions in three different experiments, analogous to the Bell case, with each involving measurements at just two times and non-invasiveness is imposed only in each separate experiment. Furthermore, even within each experiment involving two times, there are a number of different choices to made, as we shall see in more detail.
A much stronger reading of NIM is to assert that it should not make any difference if one, two or three sequential measurements are made in the same experiment.
Both of these versions of NIM permit access to the information required to determine whether MRps holds, but since MR is defined to be the conjunction of both MRps and NIM, it means that there are {\it different versions} of MR depending on how the NIM requirement is implemented. 

The purpose of this paper is to explore the consequences of these different implementations of NIM.
For convenience we will denote the stronger version of NIM consisting of sequential measurements at three times by $\NIM_{seq}$ and the weaker one, in which NIM is only satisfied in a piecewise (pw) way, by $\NIM_{pw}$. These different characterizations of NIM will be further refined as required.

\subsection{Parallels with Bell Experiments}

The LG framework for testing macrorealism was designed by analogy with Bell experiments and 
it does indeed have some genuinely close parallels. 
For example, in simple models involving a single spin system with $Q$ given in terms of the Pauli matrices by $Q = {\bf a} \cdot \sigma$, where ${\bf a}$ is a unit vector,
the correlation functions have the form $C_{12} = {\bf a (t_1)} \cdot {\bf a(t_2) }$, so are identical in form to the EPRB correlation functions and violations of the LG inequalities are easily found.
Moreover, in practice, measurements of $Q$ at two different times are typically accomplished using an ancilla (see for example Ref.\cite{Knee}),
which entangles with the state of the primary system at the first time, and the correlation function is then obtained from measurements of both system and ancilla at the second time. Thus we are really dealing with an entangled pair, just like Bell experiments.

However, the analogy fails at a number of points. As Maroney and Timpson have argued \cite{MaTi}, Bell and LG tests are not methodologically on a par since the notion of non-invasiveness typically carries some model-dependent assumptions so is difficult to motivate as a general feature, unlike local causality in Bell experiments \cite{Note3}.

This paper will focus on another key difference with the Bell case which is the question of sufficient conditions for macrorealism. The LG inequalities are necessary conditions but they are not sufficient, as has been noted by a number of authors \cite{KoBr,Cle}.
By contrast in Bell experiments, Fine's theorem \cite{Fine,HalFine,Pit,GaMer,Bus}
guarantees that the Bell \cite{Bell} or CHSH \cite{CHSH} inequalities are both necessary and sufficient conditions for the existence of an underlying probability matching the given correlation functions, and so are necessary and sufficient conditions for local realism.  This means that the Bell or CHSH inequalities are a decisive test.

The point at which Fine's theorem fails to apply to the LG framework relates to the description of the system at two moments of time. The probability $p(s_1,s_2)$ for the values $s= \pm 1$ of $Q$ at times $t_1$, $t_2$,
from which the correlation function $C_{12}$ is obtained, refers to incompatible quantities (i.e. non-commuting ones in quantum mechanics), whereas the analogous quantities in the Bell case are compatible. This is a reflection of no-signaling in the Bell case and the no-signaling conditions are a key assumption in Fine's theorem since they ensure that all the pairwise probabilities are consistent with each other. The analgous conditions do not hold in the LG framework. For example, suppose we carry out sequential measurements of $Q$ at $t_1$ and $t_2$, yielding probability $p_{12} (s_1,s_2)$ and compare with the probability $p_{23} (s_2,s_3)$ obtained by carrying out sequential measurements at $t_2$ and $t_3$. We would in general find that expected relations of the form
\beq
\sum_{s_1} p_{12} (s_1,s_2) = \sum_{s_3} p_{23} (s_2,s_3),
\label{MS}
\eeq
do not hold. This means in essence that MR can already fail at the two-time level, a feature not normally discussed in the LG framework.
This difference with the Bell case means that the LG inequalities alone are not a decisive test of MR since they could be satisfied even when MR fails.

This naturally leads to the question as to whether this difference can be rectified, i.e. do there exist conditions for MR which are decisive? 
As noted above the underlying issue in the LG framework is that, from a quantum-mechanical perspective, the probabilities $p(s_1,s_2)$ are probabilities for a pair of non-commuting observables. Quantum mechanics may still assign probabilities to such observables, but there are
a number of different ways of doing so and they often come with additional conditions.
Consequently, we will find that the difference between the LG and Bell framework can in fact be rectified, but in at least two very different ways, corresponding to different implementations of NIM.

\subsection{Necessary and Sufficient Conditions for Macrorealism}

The first way to derive necessary and sufficient conditions for MR that we shall explore is to work with the weaker form of non-invasiveness, $\NIM_{pw}$, and
stay as close as possible to the original LG framework, in which the correlation functions are determined in a number of different runs, and find a way to fill the shortfall between the LG inequalities and the requirements of Fine's theorem. Since this shortfall arises because sequential measurements will not in general satisfy no-signaling type conditions of the form Eq.(\ref{MS}), we seek another way of finding information about the system at two times which does not solely involve sequential measurements. This, we will show,
consists of doing sufficiently many experimental runs so that only compatible quantities are measured in each run, non-invasively, and then deriving LG-type inequalities for {\it two times} which determine whether or not MR holds at two times. This yields a set of two-time and three-time LG inequalities which, when the two-time LG inequalities hold, have a mathematical form identical to that of the Bell system and are therefore necessary and sufficient conditions for MR. Hence, the desired parallel with the Bell system and a decisive test for MR is achieved using an augmented set of LG inequalities measured in a judiciously chosen set of runs.  The measurements are non-invasive, by design, so this protocol is perhaps most accurately thought of as a direct test of MRps.
This approach was outlined already in Ref.\cite{HalQ} but is re-iterated here firstly, to give a very different presentation which stresses and amplifies a number of significant features and secondly, to compare with the second approach described below.

The second way is to follow the much stronger reading of NIM outlined above, $\NIM_{seq}$, and restrict to initial states and other parameter ranges so that
relationships of the form Eq.(\ref{MS}) hold for sequential measurements. Such relationships were named no-signaling in time (NSIT) conditions by Kofler and Brukner \cite{KoBr}. In particular, Clemente and Kofler \cite{Cle} proposed a scheme in which the underlying three-time probability $p_{123}(s_1,s_2,s_3) $
is determined in a single experiment by sequential measurements at all three times, subject to a set of two- and three-time NSIT conditions, similar to Eq.(\ref{MS}).
When these conditions hold, the probability $p_{123}(s_1,s_2,s_3)$ is a properly defined probability for a set of three independent variables, and
hence the set of NSIT conditions are a necessary and sufficient condition for MR.
The LG inequalities are not involved in this sort of test, but are clearly implied by the set of NSIT conditions. These conditions test a combination of NIM and MRps (and induction).
It is very different to a Bell test since it involves sequential measurement of incompatible quantities. 
This test is also of the same type as some of the ``coherence witness'' tests proposed recently \cite{Rob,Wit,Ema}.

These two possibilities clearly delineate two extremes in terms of how strongly or weakly NIM is implemented. We will also find intermediate possibilities that involve combinations of both. (We also note here a possible connnection with the so-called Wigner Leggett-Garg inequalities, which lie midway between the LG inequalities and no-signaling conditions \cite{WLG}).

The different varieties of NIM explored in this paper are clearly matters over which the experimentalist has choice and control, and therefore likewise the consequent definitions of macrorealism under test. 
Hence, it is not the purpose of this paper to promote any particular version of NIM and MR ahead of another. Rather, the purpose is simply to classify and compare different definitions of MR.

Note also that in talking about measurements which we refer to as ``non-invasive'', we have in mind a theoretical ideal situation (for example that in which only compatible quantities are measured in the same experiment, hence there is no possibility, in principle, for one measurement to disturb another). In practice experimental clumsiness is difficult to eliminate and this leaves loopholes for alternative explanations of the results \cite{deco}. See Refs.\cite{Knee,Rob} for further discussions of how this may be handled in specific experiments.

\subsection{This Paper}

We begin in Section 2 by describing the EPRB experiment in some detail. This is to assist the comparison with LG tests. We note in particular that first of all, Bell tests involve combining probabilities for incompatible quantities obtained from different experiments and secondly,
the Bell/CHSH inequalities can be satisfied in face of non-zero quantum coherence.
In Section 3 we describe some aspects of the measurement of temporal correlation functions and motivate the procedure of determining which variables to measure in each experimental run.
Tests of macrorealism involving the augmented LG inequalities are described in Section 4, and tests involving NSIT conditions are described in Section 5. Some quantum-mechanical aspects of the LG and NSIT approaches are briefly discussed in Section 6 along with a simple property of coherence witnesses.
We summarize and conclude in Section 7.

\section{The EPRB Experiment}

To fix ideas it is very useful to briefly review the EPRB experiment.
We consider a
pair of particles $A$ and $B$ in the entangled state 
\beq
| \Psi \rangle   = \frac {1} {\sqrt{2} } \left ( |\! \uparrow \rangle \otimes | \! \downarrow \rangle
- |\! \downarrow \rangle \otimes |\!  \uparrow \rangle \right),
\eeq
where $ |\! \!\uparrow \rangle $ and $ | \!\!\downarrow \rangle$  denote spins in the $z$-direction \cite{Bell,CHSH}.
Measurements are made on the spin of $A$ in directions $\av$ or $\av'$, with outcomes $s_1, s_2$ taking values $\pm 1$, and on $B$ in directions $\bv$ or $\bv'$ with outcomes $s_3,s_4$. We thus determine the four probabilities $p(s_1,s_3),  p(s_1,s_4), p(s_2,s_3), p(s_2,s_4)$. In quantum mechanics they are given by
\beq
p(s_1,s_3) = \langle \Psi | P_{s_1}^{\av} \otimes P_{s_3}^{\bv} | \Psi \rangle,
\label{prob}
\eeq
where the projection operators onto spin in direction $\av$ are defined in terms of the Pauli matrices by
\beq
P_s^\av = \half \left( 1 + s \av \cdot \sigma \right).
\eeq
These probabilities obey no-signaling conditions, of the form
\beq
\sum_{s_1} p(s_1,s_3) = p(s_3) = \sum_{s_2} p(s_2,s_3),
\label{loc}
\eeq
where $p(s_3) = \langle \Psi | P_{s_3}^{\bv}| \Psi \rangle $.
Suppose the four pairwise probabilities can be regarded as the marginals of an underlying probability $p(s_1,s_2,s_3,s_4)$, so that, for example,
\beq
p(s_1,s_3) = \sum_{s_2,s_4} p(s_1,s_2,s_3,s_4).
\eeq
If such a probability exists then the correlation functions $C_{13}, C_{14}, C_{23} $ and $C_{24}$, defined by
\beq
C_{ij} = \sum_{s_1,s_2,s_3,s_4} s_i s_j p(s_1,s_2,s_3,s_4),
\eeq
must satisfy the eight CHSH inequalities, consisting of the two relations
\beq
-2 \le C_{13} + C_{14} + C_{23} - C_{24} \le 2,
\label{CHSH}
\eeq
plus six more obtained by moving the minus sign in front of $C_{24}$ to the three other possible locations \cite{CHSH}. 
According to Fine's theorem, these eight inequalities are also a sufficient condition to guarantee the existence of an underlying probability \cite{Fine,HalFine,Pit,GaMer,Bus}. The CHSH inequalities are therefore a definitive test of local realism.

It is not hard to find quantum states for which these inequalities are violated and this has also been experimentally verified. Hence quantum theory exhibits many situations in which local realism cannot be maintained.

A number of comments can be made here, for the sake of future comparison with the Leggett-Garg situation. The measurements are carried out using four experiments,  where each experiment measures a pair of quantities which are compatible (commuting, in the quantum description), which means that each pairwise probability, $p(s_1,s_3)$ for example, is unambiguously defined and obeys  no-signaling conditions for the form Eq.(\ref{loc}).
However, the sought after
underlying probability $p(s_1,s_2,s_3,s_4)$ matching the given marginals includes some {\it incompatible} pairs. This is the essence of this sort of test -- to determine whether a set of quantities which in quantum mechanics are non-commuting nevertheless have a local description in which they may be assigned definite values.
The determination of the existence or not of this probability is carried out {\it indirectly}, by taking a series of partial snapshots of the system and then using the CHSH inequalities and Fine's theorem to determine whether the partial snapshots are consistent with an underlying notion of local realism. 

One might contemplate instead attempting to determine the underlying probability directly using sequential measurements of the incompatible variables. If the measurements in the $\av$ and $\bv$ directions were measured first, followed by measurements in the $\av'$ and $\bv'$ direction, then resulting quantum-mechanical measurement probability is
\beq
p_{1234} (s_1,s_2,s_3,s_4) =    \langle \Psi | P_{s_1}^{\av} P_{s_2}^{\av'}  P_{s_1}^{\av} \otimes  P_{s_3}^{\bv}P_{s_4}^{\bv'}  P_{s_3}^{\bv} | \Psi \rangle.
\label{QMprob}
\eeq
However, this would not yield a true probability for four independent variables
because measurement probabilities of this form for non-commuting variables do not obey the probability sum rules.
For example, summing out $s_2$ and $s_4$ gives the expected resut for $p(s_1,s_3$), Eq.(\ref{prob}), but summing out $s_1$ and $s_3$ does not give the corresponding expected result for $p(s_2,s_4)$, except perhaps for special initial states or very particular choices of the four spin vectors.
This failure of the sum rules is due to quantum interference. In simple physical terms, the first measurement disturbs the result of the second. Because of this feature we do not use sequential measurements of the non-commuting variables in the EPRB experiment to test local realism. (Although note however interesting results can be obtained if the first measurement is weak \cite{Hig}).

Note also that requiring Eq.(\ref{QMprob}) to satisfy the sum rules, i.e. requiring zero interference, is a much stronger condition than
the CHSH inequalities. This means that the CHSH inequalities 
can be satisfied and thus an underlying probability can exist even when the sum rules for Eq.(\ref{QMprob}) fail. That is, a local hidden variables model replicating the correlation functions can exist even in the face of non-zero interferences, as long as they are not too large.

\section{Measuring Temporal Correlation Functions}

We now consider the non-invasive measurement of the temporal correlation functions, by way of preparation for the augmented LG protocol in the next section.
The original LG framework envisaged the measurement of a two-time probability $p(s_1,s_2)$ from which the correlation function is obtained,
\beq
C_{12} = \sum_{s_1,s_2} s_1 s_2 \ p (s_1, s_2).
\eeq
From this probability one can also determine the averages, 
\bea
\langle Q_1 \rangle
&=&  \sum_{s_1,s_2} s_1 \ p(s_1, s_2),
\\
\langle Q^{(1)}_2 \rangle &=&  \sum_{s_1,s_2} s_2 \ p(s_1, s_2),
\eea
where we use the shorthand $Q_i$ to denote $Q(t_i)$. The superscript ${(1)}$ acknowledges the possibility that the value of $Q_2$ could be disturbed by the earlier measurement at $t_1$.
These quantities are generally not required in standard LG tests but will be utilized in the more comprehensive tests of MR considered here.
The two averages and the correlation function uniquely determine the probability, via the moment expansion,
\beq
p(s_1, s_2) = \frac {1} {4} \left( 1 + s_1 \langle Q_1 \rangle + s_2 \langle Q^{(1)}_2 \rangle
+ s_1 s_2 C_{12} \right).
\eeq
(This useful representation is described in more detail in Refs.\cite{HaYe,Kly}).

The probability $p(s_1,s_2)$ is typically determined by sequential measurements involving an ideal negative measurement at the first time, which means that there is no possibility from a macrorealistic perspective that the value of the correlation function can be explained to be the result of the disturbance produced by the first measurement. 
However, as indicated if we were to measure $\langle Q_2^{(1)} \rangle$ we would find that it {\it is} in fact disturbed by the earlier measurement, at least for some initial states, so would not be the same as the quantity $\langle Q_2 \rangle $ obtained in the absence of an earlier measurement.
This is because the experimental apparatus will obey the laws of quantum mechanics and ideal negative measurements still cause wave function collapse, even though they are non-invasive from the macrorealistic point of view \cite{Dicke}. For this reason, sequential measurements generally do not obey NSIT conditions, such as
\beq
\sum_{s_1} p_{12} (s_1,s_2) = p_2 (s_2),
\label{NSIT}
\eeq
where $p_{12} (s_1,s_2)$ denotes the probability obtained from measurements at both $t_1$ and $t_2$ and $p_2 (s_2)$ denotes the probability obtained from a measurement at $t_2$ only, with no earlier measurements.
In the NSIT protocol to be described in Section 4 in which $\NIM_{seq}$ is implemented,
Eq.(\ref{NSIT}) is quite simply enforced by restriction of the parameters of the model. However, the augmented LG protocol to be described in Section 3, in which $\NIM_{pw}$ is implemented, offers a different way of proceeding 
and we now give the background to this.

The macrorealist may have some difficulty explaining the failure of ideal negative measurements to satisfy Eq(\ref{NSIT}).
On the one hand, there is a macrorealistic argument for the non-invasiveness of ideal negative measurements, yet on the other hand, $\langle Q_2 \rangle$ can be measurably disturbed, which casts doubt on the validity of the argument that the value of $C_{12}$ cannot be explained by a classical model with disturbing measurements. 
This feature of ideal negative measurements is normally not problematic since most experiments are interested only in the correlation function, and not the value of  $\langle Q_2 \rangle$. However, in the present approach, in which we are addressing the question as to whether MR holds at the two-time level, we will also need the values of the averages of $Q$, so this feature needs to be addressed.

In the Bell case considered in the previous section, in each experimental run, measurements are made only of quantities which are compatible, namely only one spin component for each particle in each run. One then attempts to combine the results from different incompatible runs into a single probability. The Leggett-Garg case is fundamentally different in this respect in that the two-time probability refers to the probability for two non-commuting operators $\hat Q_1$, $\hat Q_2$. However, there is more similarity with the Bell case than might be immediately apparent.

From a quantum-mechanical point of view, the quantities we are interested in determining are the averages of the operators $ \hat Q_1$, $ \hat Q_2$, and their anticommutator operator,
\beq
\hat C_{12} =\half \left( \hat Q_1 \hat Q_2 + \hat Q_2 \hat Q_1 \right),
\eeq
since the correlation function is given by $C_{12} = \langle \hat C_{12} \rangle$. 
(This operator is trivially proportional to the identity in the simplest spin models, but not so for more general models).
The operator $\hat C_{12}$ has the (not immediately obvious) properties that it commutes with $\hat Q_1$ and $\hat Q_2$,
\beq
[ \hat Q_1, \hat C_{12} ] = 0 = [ \hat Q_2, \hat C_{12} ].
\eeq
This property, previously noted in Ref.\cite{HalLG2}, follows  from the fact that $\hat Q^2 = 1$.
This means that the pair $\hat Q_1$ and $\hat C_{12}$ are in fact {\it compatible} quantities, even though $\hat Q_1$ and $\hat Q_2$ are not. For short time intervals, this property reduces to the 
statement that $ \hat Q$  commutes with $(  d \hat Q / d t )^2 $.
The latter quantity is a measure of whether $\hat Q$ is about to change sign or not, in either direction. 
So although the velocity $ d \hat Q / d t$ will be disturbed by a measurement of $\hat Q$, it
is possible, perhaps surprisingly, to specify both the value of $\hat Q$ and whether it is about to change sign.


In the interests of non-invasiveness, it is then very natural
to separate the determination of $ \langle \hat Q_1 \rangle$, $\langle \hat Q_2 \rangle$ and $C_{12}$ into two separate experiments, in which  $ \langle \hat Q_1 \rangle$ and $C_{12}$ are determined in one experiment and $\langle \hat Q_2 \rangle $ is determined in a separate experiment. In that way only compatible quantities are measured in each run, as in the Bell case.

Furthermore, this sheds some light on the apparently contradictory feature of ideal negative measurements noted above. If we use an ideal negative measurement to measure only $  \langle  Q_1 \rangle$ and $C_{12}$ in a single experiment then there is no sense, even at the quantum level, that the measurement of $Q_1$ in some way disturbs the value of $C_{12}$ because we are measuring compatible quantities. (Or in other words, $C_{12}$ is insensitive to superpositions of eigenstates of $\hat Q_1$.)
However, a subsequent measurement of $ \langle Q_2 \rangle $ would be disturbed since $ Q_1 $ and $Q_2$ are incompatible. Hence it makes sense to reject the value of $\langle Q_2 \rangle$ determined as part of two sequential measurements, since it will have been disturbed by the earlier measurement, and instead measure $ \langle Q_2 \rangle $ in a different run.

Although in the above discussion we are using the quantum-mechanical notion of incompatible, this can clearly be determined operationally without recourse to quantum mechanics. By doing a number of different experiments the macrorealist could determine which sets of quantities can be measured together without disturbing each other.

Note also that we are talking about $C_{12}$ as if it was a separate quantity from $Q_1$, whereas in sequential measurements $C_{12}$ is determined by measuring $Q_1$ followed by $Q_2$. However, there are in fact measurement protocols in which $C_{12}$ can be measured directly without determining $Q_1$ or $Q_2$. For example, the ``waiting detector'' model of Ref.\cite{HalLG2} measures only whether or not $Q(t)$ changes sign during the time interval $[t_1,t_2]$, from which the correlation function $C_{12}$ is readily determined, but without determining $Q_1$ or $Q_2$. A similar protocol is described in Ref.\cite{HalLG3}, in which an ancilla registers the value of $C_{12}$ but without collapsing superposition states of $ \hat Q_1$.


The above observations indicate that it is useful to regard invasiveness as consisting of two distinct components. There is the invasiveness that would be present classically in the presence of interaction with a measuring device. This invasiveness can be avoided using an ideal negative measurement. But there is also a second type of invasiveness that arises only when incompatible quantities are measured sequentially. This is clearly a quantum effect and is not avoided by an ideal negative measurement, but can be avoided by using different experiments for incompatible quantities as proposed here.
The macrorealist can offer no understanding of the incompatibility of certain measurements, but can check for it experimentally and hence avoid it by a judicious choice of experimental runs.

In summary,
the weaker sense of outlined in the Introduction, $\NIM_{pw}$, can be implemented
by grouping the variables one wishes to measure into compatible sets and measuring only compatible variables in each experimental run. 
Macrorealistic arguments for non-invasiveness then persist to the quantum level which 
may then be upheld by experiments (subject to the caveats expressed in Section 1 about the clumsiness loophole).

\section{Necessary and Sufficient Conditions for Macroealism using Leggett-Garg Inequalities}

We now exhibit a set of necessary and sufficient conditions for macrorealism using an augmented set of Leggett-Garg inequalities and using $\NIM_{pw}$.
Following the approach described in the previous section, we carry out a set of experiments to determine the averages and second order correlation functions of $Q(t)$ at three times, by measuring only compatible quantities in each experimental run and using ideal negative measurements. 
There are numerous ways to group the compatible quantities. A convenient choice is to do four experiments in which we measure
$\langle Q_1 \rangle$ and $C_{12}$ in the first run,  $ \langle Q_2 \rangle$ and $C_{23}$ in the second, $C_{13}$ in the third and $ \langle Q_3 \rangle $ in the fourth. (With a different type of measurement one could also consider combining the last two into a single experiment since $C_{13}$ and $\langle Q_3 \rangle $ are compatible).

Macrorealism is the question as to whether or not there exists an underlying probability $p(s_1,s_2,s_3)$ matching the six moments measured in the way described above. We build this up in three steps.

First, the single time probabilties are given by
\beq
p(s_i) = \frac{1}{2} \left( 1 + s_i \langle Q_i \rangle \right),
\eeq
for $i=1,2,3$, and these are non-negative by construction. Second, there are three two-time probabilties, given by
\beq
p(s_i,s_j) = \frac{1}{4} \left( 1 + s_i \langle Q_i \rangle + s_j \langle Q_j \rangle + s_i s_j C_{ij} \right),
\label{qmom}
\eeq
where $ij =12,13,23$. Because the averages $ \langle Q_i \rangle$ are measured in such a way that there can be no disturbance from an earlier measurement,
these two-time probabilities will obey all compatibility conditions of the form,
\beq
\sum_{s_i} p(s_i,s_j) = p(s_j) = \sum_{s_k} p(s_j,s_k).
\label{comp}
\eeq
These relations are of course mathematically identical to the NSIT conditions, Eq.(\ref{NSIT}), but there is no sense in which they indicate the absence of ``signaling'', since the two-time probabilities are assembled indirectly from different experiments, not measured sequentially in a single experiment.
Instead these relations are simply the compatibility relations between the two-time probabilities that are required for Fine's theorem to apply.

In a macrorealist theory in which the averages and correlation function are non-invasively measured, the two-time probabilities Eq.(\ref{qmom}) are guaranteed to be non-negative. This follows very easily from a simple argument similar to the derivation of the LG inequalities: we have
\beq
(1 + s_i Q_i) ( 1 + s_j Q_j ) \ge 0,
\eeq
and averaging this, we obtain
\beq
1 + s_i \langle Q_i \rangle   + s_j   \langle Q_j \rangle   + s_i s_j C_{ij}  \ge 0.
\label{LG2}
\eeq
These twelve conditions, which we will call {\it two-time Leggett-Garg inequalities} are necessary conditions for macrorealism at the two-time level. They are also sufficient because if satisfied, the left-hand side of Eq.(\ref{LG2}), multiplied by $\frac{1}{4}$, are precisely the probabilities 
Eq.(\ref{qmom}) matching the given averages and correlation functions.

Finally, the most general possible form of the desired three-time probability is
\bea
p(s_1,s_2,s_3) &=& \frac{1}{8} \left( 1 + s_1 \langle Q_1 \rangle + s_2 \langle Q_2 \rangle +s_3 \langle Q_3 \rangle 
\right.
\nonumber \\
&+&  \left. s_1 s_2 C_{12} + s_2 s_3 C_{23} + s_1 s_3 C_{13} + s_1 s_2 s_3 {D}
\right) .
\label{p123}
\eea
It involves a coefficient $D$, essentially the triple correlator, which is {\it not} measured in the experiment. The question is whether there is any possible value of $D$ for which
\beq
p(s_1,s_2,s_3) \ge 0.
\label{p0}
\eeq
Fine's theorem guarantees that this is indeed possible under the following conditions: the twelve two-time LG inequalities Eq.(\ref{LG2}) hold; the compatibility conditions Eq.(\ref{comp}) hold; and the four three-time LG inequalities, Eqs.(\ref{LG1})-(\ref{LG4}) hold.

The proof of this result is spelled out in detail in Ref.\cite{HalFine}. However, it is easily seen as follows.
In the inequality Eq.(\ref{p0}),
the four values of $s_1,s_2,s_3$ for which $s_ 1 s_2 s_3 = - 1$ yield four upper bounds on $D$.
Similarly the four values of $s_1,s_2,s_3$ for which $s_ 1 s_2 s_3 =  1$ yield four lower bounds on $D$. Hence there exists a $D$ for which Eq.(\ref{p0}) holds as long as the four lower bounds are less than the four upper bounds. These yields sixteen relations, which consist of precisely the twelve two-time LG inequalities and the four three-time LG inequalities.

The new feature in this protocol, compared to standard LG tests, are the twelve two-time LG inequalities, Eq.(\ref{LG2}). It is these that fill the shortfall in the usual three-time LG inequalities and lead to conditions for MR which are both necessary and sufficient.

Concisely summarized, the protocol just described tests a specific definition of MR, consisting three sets of two-time LG inequalities, one set of three-time inequalities, together with induction and piecewise non-invasive measurability. This is definition of MR is arguably the weakest one possible, and we write,
\beq
\MR_{weak} = \NIM_{pw} \wedge \LG_{12} \wedge  \LG_{23} \wedge \ LG_{13}  \wedge \LG_{123} \wedge \Ind.
\eeq
Like the Bell and CHSH inequalities, it may be satisfied in the face of non-zero interferences, as long as they are not too large.

The twelve two-time and four three-time LG inequalities can be readily simplified by a particular choice of initial state \cite{EmINRM,SimpLG,Ema}. 
Suppose that we choose the initial state of the system to be an eigenstate of $\hat Q_1$ at time $t_1$, with eigenvalue $+1$. Then $C_{12} = \langle Q_2 \rangle $ and $C_{13} = \langle Q_3 \rangle$.
The four three-time LG inequalities Eq.(\ref{LG1})-(\ref{LG4}), then read,
\bea
1 +  \langle Q_2 \rangle + \langle Q_3 \rangle + C_{23} &\ge& 0, 
\label{D1}\\
1 -  \langle Q_2 \rangle - \langle Q_3 \rangle + C_{23} &\ge& 0, \\
1 +  \langle Q_2 \rangle - \langle Q_3 \rangle - C_{23} &\ge& 0, \\
1 -  \langle Q_2 \rangle + \langle Q_3 \rangle - C_{23} &\ge& 0,
\label{D4}
\eea
which therefore coincide with four of the twelve two-time LG inequalities. The remaining eight two-time LG inequalities
consist of trivially satisfied conditions of the form $| \langle Q_i \rangle | \le 1 $. Hence in this simplied situation the four inequalities Eq.(\ref{D1})-(\ref{D4}) are necessary and sufficient conditions for $\MR_{weak}$. Inequalities of this general form have been tested experimentally \cite{EmINRM,SimpLG,Ema}.

The above protocol is readily extended to the four-time situation, for which we find,
\beq
\MR_{weak} = \NIM_{pw} \wedge \LG_{12} \wedge  \LG_{23} \wedge \ LG_{34} \wedge LG_{14}  \wedge \LG_{1234} \wedge \Ind.
\eeq
That is, there are four two-time LG inequalities together with the eight four-time LG inequalities, which have the form
\beq
-2 \le C_{12} + C_{23} + C_{34} - C_{14} \le 2,
\label{LG4}
\eeq
plus the three more pairs of inequalities obtained by moving the minus sign to the other three possible positions.

\section{Necessary and Sufficient Conditions for Macrorealism using No-Signaling in Time}

We now review very different conditions for macrorealism which make use of no-signaling in time conditions and do not involve the LG inequalities at all. The most comprehensive version of this approach is that of Clemente and Kofler \cite{Cle} which is followed here.  (Coherence witness conditions \cite{Rob,Wit,Ema} are simpler examples of this approach and conditions similar to the ones that follow have been given by Maroney and Timpson \cite{MaTi}).
They suppose that the system is measured using sequential measurements at three times, with all measurements done in the same experiment and then conditions are imposed to ensure that these measurements are non-invasive, hence we are working with $\NIM_{seq}$. 
This procedure accesses the underlying three-time probability $p(s_1,s_2,s_3)$ directly but the nature of the measurements means that the procedure works only under conditions
considerably stricter than those required in the augmented LG tests.

In the face of potentially invasive sequential measurements, the most general possible form for the three-time probability is
\bea
p_{123} (s_1,s_2,s_3) &=& \frac{1}{8} \left( 1 + s_1 \langle Q_1 \rangle + s_2 \langle Q_2^{(1)} \rangle +s_3 \langle Q_3^{(12)} \rangle 
\right.
\nonumber \\
&+&  \left. s_1 s_2 C_{12} + s_2 s_3 C_{23}^{(1)}  + s_1 s_3 C_{13}^{(2)} + s_1 s_2 s_3 {D}
\right). 
\label{pm}
\eea
Here, the superscripts again acknowledge that the values of averages and correlation functions can depend on whether earlier or intermediate measurements are made. So for example, 
$ \langle Q_3^{(12)} \rangle  $ can depend on whether measurements were made at both $t_1$ and $t_2$, and $C_{13}^{(2)} $ can depend on whether a measurement is made at the intermediate time $t_2$. In contrast to the three-time probability discussed in the LG framework, Eq.(\ref{p123}), here, the triple correlator $D$ is determined by the measurement process.
We assume induction throughout so there is no possibility of dependence on later measurements \cite{Note1}.

Eq.(\ref{pm}) is non-negative by definition since it is a measurement probability. However, because of the possible dependencies of its components on the context of the measurement, it is not the probability of an independent set of variables, so is not yet the sought after description of macrorealism we seek. Clemente and Kofler therefore imposed a set of NSIT conditions at two and three times to ensure this.

We consider the related measurement probabilities in which measurements are made at only two times, or just one time:
\bea
p_{13} (s_1,s_3) &=&\frac{1}{4} \left( 1 + s_1 \langle Q_1 \rangle +s_3 \langle Q_3^{(1)} \rangle 
+ s_1 s_3 C_{13} \right),
\\
p_{23} (s_2,s_3) &=&\frac{1}{4} \left( 1 + s_2 \langle Q_2 \rangle +s_3 \langle Q_3^{(2)} \rangle 
+ s_2 s_3 C_{23} \right),
\\
p_{12} (s_1,s_2) &=&\frac{1}{4} \left( 1 + s_1 \langle Q_1 \rangle +s_2 \langle Q_2^{(1)} \rangle 
+ s_1 s_2 C_{12} \right),
\\
p_3 (s_2) &=& \half \left( 1 +\langle Q_3 \rangle  \right).
\eea
Clemente and Kofler then impose the NSIT condition
\beq
\sum_{s_2} p_{23} (s_2,s_3) = p_3( s_3)
\label{NSIT23}
\eeq
conveniently denoted $\NSIT_{(2)3}$,
which implies that $\langle Q_3^{(2)} \rangle  = \langle Q_3 \rangle $. The NSIT condition
\beq
\sum_{s_1} p_{123} (s_1,s_2,s_3) =p_{23} (s_2,s_3),
\eeq
which we denote $\NSIT_{(1)23}$,
implies that $C_{23}^{(1)} = C_{23}$, $\langle Q_2^{(1)} \rangle =\langle Q_2 \rangle $, and
$\langle Q_3^{(12)} \rangle = \langle Q_3^{(2)} \rangle  $ (which therefore equals
$\langle Q_3 \rangle $).
Finally, the NSIT condition
\beq
\sum_{s_2} p_{123} (s_1,s_2,s_3) = p_{13} (s_1,s_3) 
\eeq
which we denote $\NSIT_{1(2)3}$,
implies $C_{13}^{(1)} = C_{13}$ and $\langle Q_3^{(12)} \rangle = \langle Q_3^{(2)} \rangle  $ (and so they are both equal to $\langle Q_3 \rangle $).
These three NSIT conditions therefore establish that all averages and correlation functions take values independent of whether earlier measurements were performed, and the three time probability may then be written:
\bea
p_{123} (s_1,s_2,s_3) &=& \frac{1}{8} \left( 1 + s_1 \langle Q_1 \rangle + s_2 \langle Q_2 \rangle +s_3 \langle Q_3 \rangle 
\right.
\nonumber \\
&+&  \left. s_1 s_2 C_{12} + s_2 s_3 C_{23}  + s_1 s_3 C_{13} + s_1 s_2 s_3 {D}
\right). 
\label{pm2}
\eea
Hence this combination of NSIT conditions are necessary and sufficient conditions for a variety of macrorealism that is clearly stronger than that described in the previous section, and we write,
\beq
\MR_{strong} = \NSIT_{(2)3} \wedge \NSIT_{(1)23} \wedge \NSIT_{1(2)3} \wedge \Ind.
\eeq
From the quantum-mechanical point of view, the NSIT conditions can only hold if the interferences are zero.

There are other combinations of NSIT conditions which achieve the same result \cite{Cle}. An extension to the four-time case is presumably possible. It will not be described here, but the moment expansion for four dichomotic variables, given in Ref.\cite{HaYe}, is a useful starting point.

In contrast to the LG case, where the measurements are non-invasive by design, the sequential measurements used in these NSIT conditions are invasive in general. In any experimental test it is therefore necessary to adjust the initial state and measurement times (and perhaps other parameters too) to ensure that the NSIT conditions are satisfied. This is why this definition of MR appears to involve far more restrictive conditions than in the augmented LG case, i.e. equalities, rather than inequalities \cite{Cle}. The NSIT conditions are primarily measures of NIM for sequential measurements, whereas NIM is already taken to be satisfied, by design, in the augmented LG case.
Of course, the values of the averages and correlation functions in Eq.(\ref{pm2}) must be the same as those determined in the augmented LG protocol, in Eq.(\ref{p123}), but the conditions under which they can be determined are different in each case: in Eq.(\ref{pm2}) they can be determined only if the equalities consisting of the NSIT conditions hold, whereas in Eq.(\ref{p123}), no such restrictions are required.

The two different types of protocols described in this and the last section are not the only possibilities and clearly delineate the two extremes. A third, intermediate option naturally arises, which is to do three experiments with sequential measurements made at only two times in each case, and then require 
that the three measured two-time probabilities all satisfy two-time NSIT conditions, of the form, Eq.(\ref{NSIT23}); in addition, we then require that the three-time LG inequalities are satisfied. This therefore tests the following version of MR:
\beq
\MR_{int} = \NSIT_{(1)2} \wedge \NSIT_{(1)3} \wedge \NSIT_{(2)3} \wedge \LG_{123} \wedge \Ind
\eeq
Like the augmented LG protocol, it stays close to the spirit of the original LG framework and clearly supplies necessary and sufficient conditions for macrorealism. It requires zero coherence at the two-time level but allows non-zero coherences at the three-time level, as long as they are suitably bounded. This protocol readily extends to the four times, analagous to the augmented LG case for four times.

All the two-time NSIT conditions can be satisfied quite easily, in a quantum-mechanical description,
by choosing an initial state such at $\langle \hat Q(t) \rangle = 0$ at all three times. Also, 
$ \NSIT_{(1)2} $ and $\NSIT_{(1)3} $ can be satisfied by choosing an initial state at $t_1$ diagonal in $\hat Q_1$.
Furthermore, in practice, the NSIT condtitions in $\MR_{int}$ will only be satisfied approximately and it is then necessary to develop extended forms of the LG inequalities appropriate to the case in which there is some signaling. This extension has been carried out by Dzhafarov and Kujala \cite{DzKu} (and is also briefly reviewed in Ref.\cite{HalLG2}).

\section{NSIT vs Two-Time LG Inequalities in a Quantum-Mechanical Description}

In a quantum-mechanical description, a direct comparison may be made between the NSIT conditions and LG inequalities for two-times using explicit measurement formulae.
The probability for two sequential projective measurements at times $t_1, t_2$ is,
\beq
p(s_1, s_2) = {\rm Tr} \left( P_{s_2} (t_2) P_{s_1} (t_1) \rho P_{s_1} (t_1) \right),
\label{2time}
\eeq
where the projection operators $P_s (t) $ are defined by $P_s (t) = e^{iHt} P_s  e^{ -i H t } $ and
\beq
P_s = \half \left( 1 + s \hat Q \right).
\eeq
By contrast, the two-time LG inequalities Eq.(\ref{LG2}) correspond in quantum mechanics to the quantities,
\beq
q(s_1, s_2) = \half {\rm Tr} \left( \left( P_{s_2} (t_2) P_{s_1} (t_1) +  P_{s_1} (t_1) P_{s_2} (t_2) \right) \rho  \right),
\label{quasi}
\eeq
which may also be written in the form Eq.(\ref{qmom}). Eq.(\ref{quasi}) may be measured either by measuring the averages and correlation function in various runs, as described, or, as argued in Ref.\cite{HalQ}, more directly, using sequential measurements in which the first one is a weak measurement \cite{weak0}.
Eqs.(\ref{2time}) and (\ref{quasi}) have the same correlation function \cite{Fri} and same $\langle \hat Q_1 \rangle$, but differ in the average of $\hat Q (t)$ at the second time.
The sequential measurement probability Eq.(\ref{2time}) does not satisfy the NSIT conditions Eq(\ref{NSIT}) in general. By contrast, Eq.(\ref{quasi}) formally satisfies NSIT, but can be negative \cite{Note}.

The relation between these two measurement formulae is given by
\beq
p(s_1,s_2) = q(s_1,s_2) 
+ \frac{1}{8}  \langle [ \hat Q(t_1), \hat Q(t_2)] \hat Q(t_1) \rangle s_2 .
\eeq
The extra term on the right-hand side, which vanishes for commuting measurements, represents interferences (as shown more explicitly in Ref.\cite{HalQ}). If we impose NSIT on $p(s_1,s_2)$ this clearly implies that the interference term is zero and hence that $p(s_1,s_2) = q(s_1,s_2)$. This also means that $q(s_1,s_2) \ge 0 $, which is equivalent to the two-time LG inequalities, Eq.(\ref{LG2}).

However, the converse is not true: $q(s_1,s_2) \ge 0 $ clearly does not imply NSIT for $p(s_1,s_2)$.
Furthermore,
since $p(s_1,s_2)$ is always non-negative, the two-time LG inequalities $ q(s_1, s_2) \ge 0 $ will be satisfied if the interference term is bounded:
\beq
\frac{1}{8}  \left| \langle [ \hat Q(t_1), \hat Q(t_2)] \hat Q(t_1) \rangle \right|   \le p(s_1,s_2)
\eeq
This confirms in this case the general story described earlier: NSIT conditions require zero interference but the LG inequalities, like the CHSH case, require only bounded interference. 

NSIT for $p(s_1,s_2)$ and $q(s_1,s_2) \ge 0 $ are both conditions for MR at two times but they are different types of conditions. NIM is assumed to hold already in the measurement of $q(s_1,s_2)$ and $q(s_1,s_2) \ge 0 $ is therefore a direct measure of MRps. By contrast,
NSIT for $p(s_1,s_2)$ measures a combination of NIM and MRps, without being able to distinguish between them. 

Note also that some of these relations between NSIT and the two-time LG inequalities are specific to the (most commonly studied) case in which measurements are made of the dichotomic variable $\hat Q$. However, for Hilbert spaces of dimension three or more, one can consider ``degeneracy-breaking'' measurements described by one-dimensional projections $P_n$, where 
$n = 1, 2 \cdots {\rm dim} {\cal H} $ and construct the two-time measurement probability $p(n_1,n_2)$ and associated quasi-probability $q(n_1,n_2)$, and from there construct the two-time correlation functions of $Q$.
This possibility arises in some of the coherence witness measures recently studied (see for example Ref.\cite{Ema}). However, the relationship between NSIT conditions for $p(n_1,n_2)$ and $q(n_1,n_2) \ge 0$ is then not as simple as the case described above. For example, the NSIT condition
\beq
p_2 (n_2) = \sum_{n_1} p_{12} (n_1,n_2)
\label{NSITn}
\eeq
has no immediate logical relation  to the analogous relation for $p(s_1,s_2)$. This is related to the fact that $p(s_1,s_2)$ need satisfy only one probability sum rule in order to be well-defined but for $p(n_1,n_2)$ there is more than one type of probability sum rule. A consequence of this is that $q(n_1,n_2)$ can in fact be negative but Eq.(\ref{NSITn}) can still be satisfied. These interesting possibilities, which are best understood from the perspective of the consistent histories approach to quantum mechanics \cite{GH1,Gri1,Gri3,Omn1,Omn3},
will be pursued in more detail elsewhere.

With these explicit formulae in hand, there is also a connection to coherence witness conditions \cite{Rob,Wit,Ema}.
One can define a witness $W(s_2)$ measuring the degree to which NSIT is violated:
\beq
W(s_2) = \left|  \sum_{s_1} p_{12} (s_1, s_2) - p_2 (s_2) \right|.
\eeq
This is easily seen to be proportional to the interference term,
\beq
W(s_2) = \frac{1}{4}  \left| \langle [ \hat Q(t_1), \hat Q(t_2)] \hat Q(t_1) \rangle \right|.
\eeq
There is therefore a simple relation between the degree of violation of NSIT and the two-time LG inequalities. If the witness is bounded according to
\beq
\half W(s_2)  \le p(s_1,s_2),
\eeq
then $q(s_1, s_2) \ge 0 $. Hence witness conditions, which are usually used to check NSIT, can also be used to check the two-time LG inequalities. This result is also in keeping with the observation in Ref.\cite{WLG} that violations of NSIT have to reach a threshold value before the  LG inequalities are violated.

\section{Summary and Conclusion}

The purpose of this paper was to elucidate and compare two very different sets of necessary and sufficient conditions for macrorealism which differ in the way in which they implement the notion of non-invasive measurability. In the first, the weaker form, measurements are made using a number of different experiments in which only compatible quantities are measured in each run and then the underlying probability, when it exists, is assembled indirectly. The measurements are non-invasive in a piecewise way, denoted $\NIM_{pw}$.
The probability then exists provided that a set of two and three time LG inequalities  hold (and also induction). This leads to a weak notion of macrorealism, which we write,
\beq
\MR_{weak} = \NIM_{pw} \wedge \LG_{12} \wedge  \LG_{23} \wedge \ LG_{13}  \wedge \LG_{123} \wedge \Ind.
\eeq
In the second, stronger form, proposed by Clemente and Kofler \cite{Cle}, macrorealism is defined by a series of NSIT conditions for sequential measurements in which all three measurements, of incompatible quantities, are made in the same experiment, together with induction:
\beq
\MR_{strong} = \NSIT_{(2)3} \wedge \NSIT_{(1)23} \wedge \NSIT_{1(2)3} \wedge \Ind.
\eeq
An intermediate notion of MR also naturally arises, in which there are three pairwise experiments with NSIT satisfied for each pair, with all correlation functions required to satisfy the three-time LG inequalities (and induction):
\beq
\MR_{int} = \NSIT_{(1)2} \wedge \NSIT_{(1)3} \wedge \NSIT_{(2)3} \wedge \LG_{123} \wedge \Ind
\eeq
These three conditions have a clear logical connection,
\beq
\MR_{strong} \implies \MR_{int} \implies \MR_{weak},
\eeq
but the converse implications clearly do not hold.
The relation between the NSIT conditions at two times and the two-time LG inequalities was spelled out explicitly in the quantum-mechanical analysis in Section 6. We also noted a relation between the two-time LG inequalities and the degree of violation of coherence witness conditions, offering a useful way of checking the two-time LG inequalities.

$\MR_{strong}$ is primarily a measure of non-invasiveness and in the quantum case is satisfied only when the interferences are zero, so is essentially the same type of condition as a number of coherence witness conditions.
By contrast, $\MR_{weak}$ allows non-zero interferences. The measurements are non-invasive by design and hence $\MR_{weak}$ is in effect a direct test of MRps. 
Both of these types of macrorealism have been discussed and tested, at least in part, in a number of previous works.
The purpose of the present work has been to make clear that these are different notions of macrorealism, due to the different ways in which NIM is implemented,
although each clearly of interest to explore and test. 

From the perpsective of the consistent histories approach to quantum mechanics \cite{GH1,Gri1,Gri3,Omn1,Omn3}, these different notions of macrorealism correspond to the fact that there exists a hierarchy of classicality conditions. This is utilized and explored in Ref.\cite{HalMC}.

For all of the protocols described in this paper, it would clearly be of interest to check experimentally a full set of necessary and sufficient conditions for macrorealism. This should not be difficult to accomplish with a modest extension of recent experiments: two and three-time LG inequalities have been tested in many different experiments, and likewise two-time NSIT conditions. What is required is an experiment which tests the appropriate combination of such conditions.

\section{Acknowledgements}

I am very grateful Clive Emary, George Knee, Johannes Kofler, Owen Maroney and James Yearsley for many useful discussions and email exchanges about the Leggett-Garg inequalities.

\bibliography{apssamp}

\begin{thebibliography}{10}
\bibitem{Note} Note that in Ref.\cite{HalQ}, Eq.(\ref{quasi}) was referred to as a ``quasi-probability''. This nomenaclature is not
used here in order to avoid any misleading associations with unorthodox interpretations of negative probabilities (which were certainly not
intended in Ref.\cite{HalQ}).
\end{thebibliography}

\end{document}